\documentstyle[12pt,epsfig,epsf]{article}

\newcommand{\institute}[1]{\parbox{16cm}{%
\centering\normalsize \sl #1}}

\title{The PMS Project: Poor Man's Supercomputer}
\author{F. Csikor$^1$, Z. Fodor$^1$, P. Heged\" us$^1$, 
V.K. Horv\'ath$^2$, S.D. Katz$^1$, A. Pir\'oth$^1$\\
\institute{
$^{(1)}$ Institute for Theoretical Physics and 
$^{(2)}$ Department of Biological Physics,\\
 E\"otv\"os University,
   P\'azm\'any P. s\'et\'any 1/A, 1117 Budapest, Hungary}
}
\date{31 December, 1999}

\begin{document}
\psfull

%\newfont{\hvfont}{cmtt9 scaled\magstephalf} 
%\newfont{\hvfont}{cmtt9 scaled\magstep1} 

\maketitle
%\vspace{2cm}

\begin{abstract}
We briefly describe the Poor Man's Supercomputer (PMS)
project carried out
at E\"otv\"os University, Budapest. 
The goal was to construct a cost effective,
scalable, fast parallel computer
to perform numerical calculations of physical problems 
that can be implemented on a lattice with nearest neighbour 
interactions.
To this end we developed the PMS architecture using
PC components and designed a special, low cost communication hardware
and the driver software for Linux OS.
Our first implementation of PMS includes 32 nodes (PMS1).
The performance of PMS1 was tested by
 Lattice Gauge Theory simulations.
Using pure SU(3) gauge theory or the bosonic MSSM
on  PMS1  we obtained  3\$/Mflop and 0.45\$Mflop 
price-to-sustained performance ratio for double and single precision 
operations, respectively.
The design of the special hardware and the communication driver
are freely available upon request for non-profit organizations.

\end{abstract}

\section{Introduction}
Our purpose was to build a high performance supercomputer from PC elements.
We use PCs for two reasons. They have excellent cost/performance ratios 
\cite{pricewatch} and
 can easily be upgraded when faster motherboards and CPUs will be available.
The PMS project
started in 1998, and the machine is now ready for physical calculations.
Our first PMS machine (PMS1) consists of 32 PCs arranged in a three-dimensional 
$2 \times 4 \times4$ mesh. Each node has two special
communication cards providing fast communication through flat cables 
to the six neighbours.
This gives a much better performance than simple Ethernet link.

Since the machine is built from PCs, the latest versions of all programming
languages (such as C and Fortran) can be used for coding. Writing applications is
straightforward. One only has to keep in mind the 3-dimensional mesh 
structure of the
machine; no further deep understanding of how the communication works is
required. There
are some routines written in C that make communication of data between the
nodes easy. The machine works in Single Instruction Multiple Data (SIMD) mode: all
processors execute the same program, while the data they work on may differ.

Nowadays double precision floating point arithmetic is necessary for accurate
results. PMS offers this precision since the  processors have double
precision Floating Point Units (FPUs). In cases when single precision is
enough, the special MMX instruction set of AMD K6-2 processors can be used. 
This provides a much higher performance, in principle 8 times higher, 
than the standard double precision mode.

The following sections describe the hardware and software architectures
of PMS. We first give a short overview of the machine and then describe 
the hardware and the software in more detail. Some performance results are 
also presented, and an outlook is given.

\section{Overview}
\begin{figure}
  \centerline{
    \epsfig{file=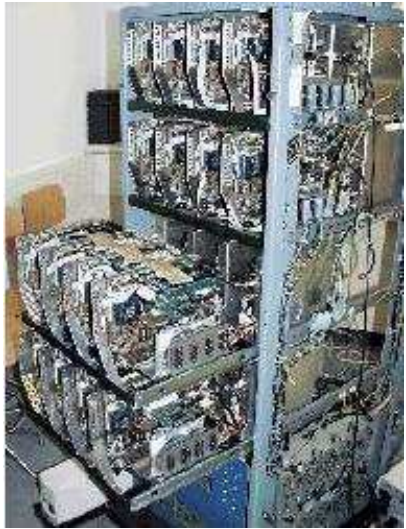,height=7cm}
  }
  \caption[hl]{The PMS1 rack and the trays.
  \label{pms_full}
  }
\end{figure}

The nodes in PMS are based on PC components. In our first PMS machine (PMS1)
the rack containes four trays (see Fig. \ref{pms_full}).
These trays  hold eight nodes in two rows.
Each node is powered by its own standard PC power supply located at the
bottom of the rack. 
Appropriate air cooling by ventilation is provided for.
Notice, that rack mounting is not
required for the present communication speed of PMS1.
 Even PCs within their own cases can be used.

 Each node in PMS is an almost complete PC.
In PMS1 the configuration of a node consists of a 
100MHz motherboard (SOYO SY-5EHM), 
a single 450MHz  AMD K6-II processor,
128Mbyte (7nsec) SDRAM, 
10Mbit Ethernet card, 
and a hard disk of caoacity 2.1 Gbyte.

The nominal speed of each node is 225 Mflop, since floating point
operations of the AMD processor require two clock cycles. The price of one
node was approximately \$350 (for the price of such a computer see
\cite{pricewatch}). The additional cost for the communication cards (see
later) is \$40 for each node. So the price/performance ratio is reasonably
low.

PMS uses special hardware  for communication (PMS CH)
to make high speed parallel
calculations possible.   
The basic idea behind the hardware is that the PMS CH  provides direct connection 
of each node to its nearest neighbours (NN). 
The PMS CH handles both polled port (PP) IO operations and direct memory access (DMA)
between two selected nodes.

The PP IO mode is quite  straightforward; 
the sender node sends a word and a flag indicating that
a word is being sent. The sender holds the data until the receiver node sends an 
acknowledgement signal after reading the data. 
In case of the DMA the problem is  synchronization
between the sender and the receiver. Typical usage of DMA assumes data transfer between a PC
and an another device. This device is usually either a slave or a bus master with 
appropriate circuits. 
If DMA is used to transfer data between two PCs then one has to deal with
synchronization between two DMA chips. Both problems are solved by using 
First In First Out (FIFO) data buffers for read(receive) and write(send) 
operations and appropriate hardware hand-shaking circuits.

Prior to data transfer the direction of the data flow must be decided.
Also, using DMA transfer receiving and sending procedures must be 
initiated in a synchronized fashion since long delays in DMA
may lead to system crash. The PC architecture limits this synchronization
to 2-30 bus clocks since both DMA and the interrupt have an 
invoking latency (which is not defined by the bus protocol and may depend on the
chip-set applied on the motherboard).

Programming  the PMS CH is a fairly simple task. It can be done under 
 Linux and 32-bit (extended) DOS operating systems. 
All low-level device drivers are written
in C and the programmer may use all kinds of commercial, share-ware or free-ware
compilers. Using the communication drivers requires the knowledge of only a 
few functions.
A simple programming example for Linux users can be found in the Appendix.

\begin{figure}
  \centerline{
    \epsfig{file=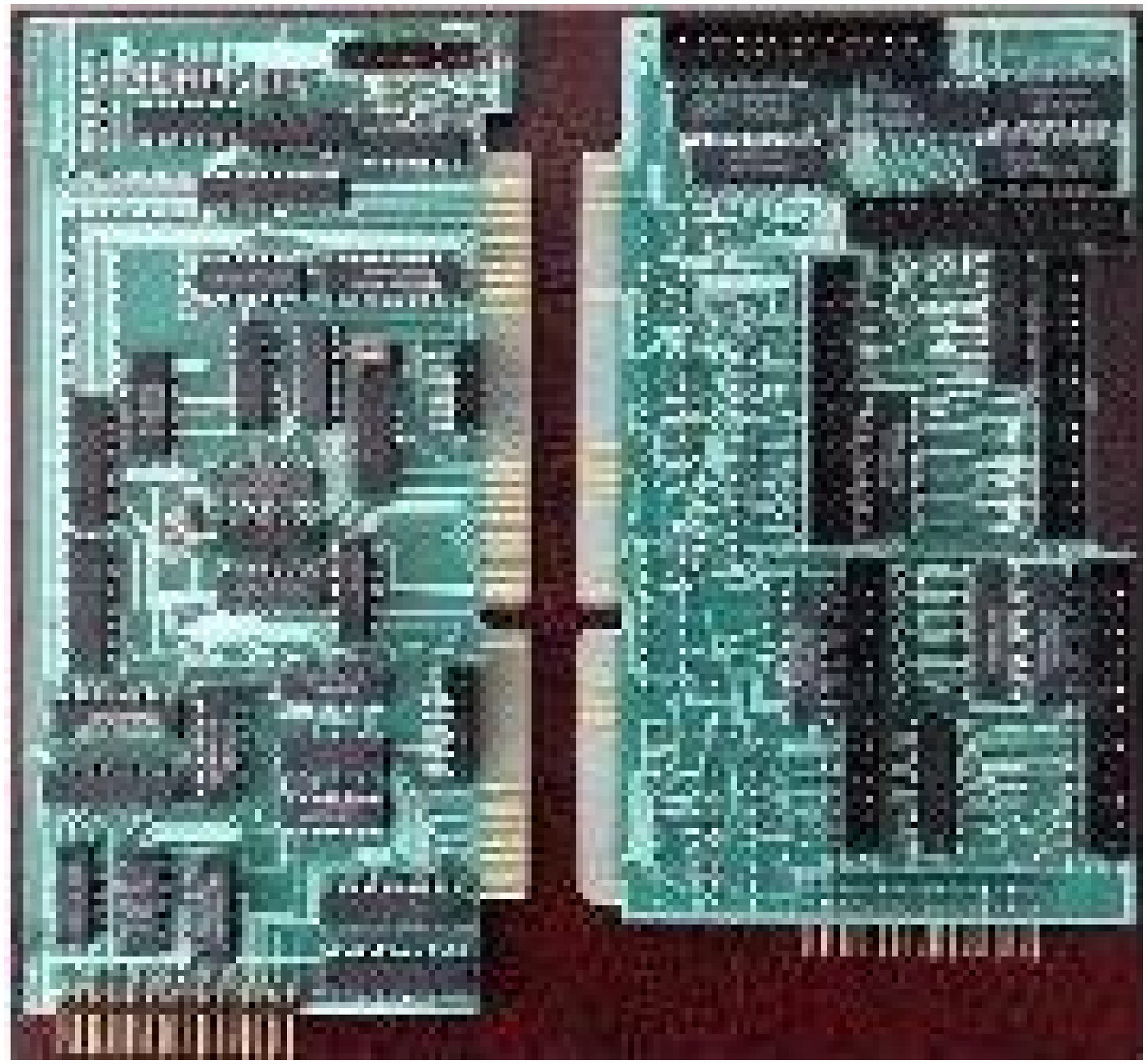,height=7cm}
     \epsfig{file=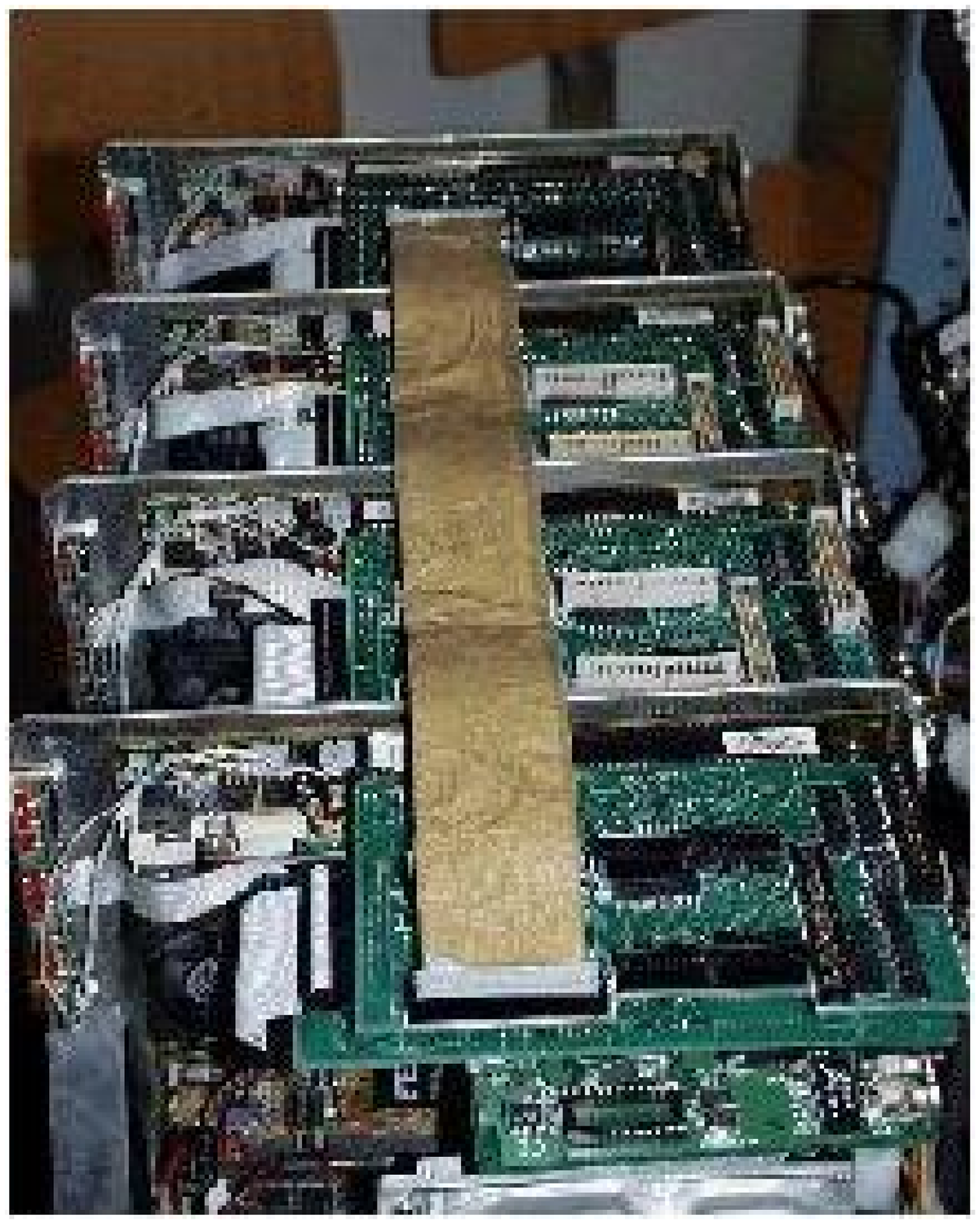,height=7cm}
  }
  \caption[a]{a, The PMS CPU and Relay cards. b, Nodes connected by flat cables in one direction.
  \label{pms_ISA} 
  }
\end{figure}

The PMS architecture does not define the bus used for the PMS CH.
In PMS1 we have used the simplest ISA bus (see Fig. \ref{pms_ISA}).
Implementation of  the PMS CH for other, faster buses is under development.

\begin{figure}
  \centerline{\epsfig{file=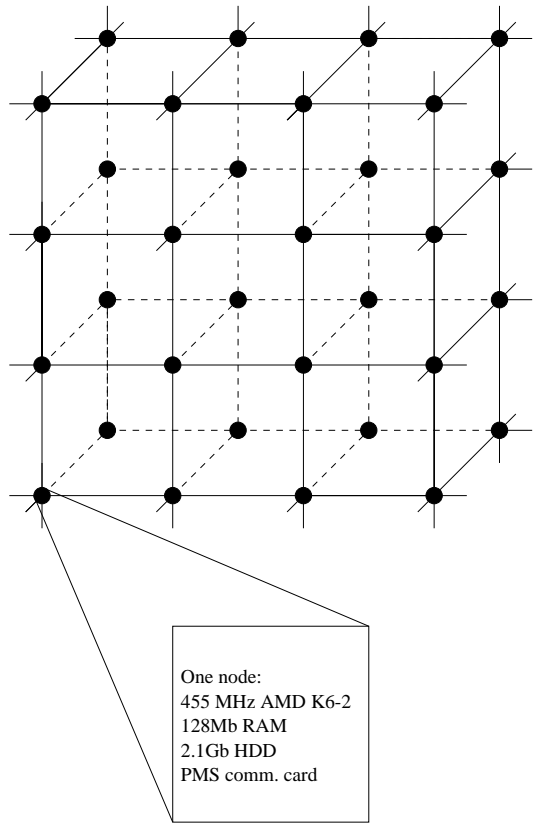,width=7cm}}
  \caption[h]{The PMS cluster \label{pms_grid}}
\end{figure}

Our first implementation of the PMS CH includes two plug-in ISA cards, 
the PMS CPU card and the PMS Relay card. 
The nodes are arranged in a $2 \times 4 \times4$  mesh as shown in Fig. 
\ref{pms_grid}.
In each node both the  PMS CPU and the PMS Relay cards are installed
providing a fast communication to the six NNs.
At the boundaries periodic boundary conditions are realized as indicated in 
Fig. \ref{pms_grid} where the links at the boundaries correspond to the ones on the other sides.
This determines the hardware architecture of the machine, which is
similar to that of the APE machines \cite{APE}.

 Debian Linux 2.1 is installed on each node. After turning the power on
each node boots from its own hard-disk. All nodes can be accessed through the
Ethernet network. There is a main computer that controls the whole cluster.
Some simple Linux scripts have been written to copy the executable
program code
and the appropriate data to and from the nodes and to execute the programs.
Collecting the results is also managed by a simple Linux script.
In principle, Ethernet connection could be used for data transfer between the nodes during simulations. However, 
this turns out to be too slow in most cases. One major reason for this is
that any data transfer between two machines makes the whole network busy.
Since building up the Ethernet connection is quite slow even for two
computers, the Ethernet connection is not satisfactory. The  
theoretical maximum 
of the Ethernet connection speed is 1 Mbyte/sec.
This bottleneck could be avoided by using switch boxes, however efficient 
switch boxes are far too expensive and would dominate  the price of the
entire machine (cf. Fig. \ref{compare} and the corresponding discussion 
in the text below.)

The special communication cards --described in more detail in the next
section-- provide faster communication between adjacent nodes. However, this
makes the machine applicable to local problems only, where
communication only between the neighbouring nodes is necessary. In PMS1 
the speed of 
communication through these cards --limited essentially by the ISA bus speed-- 
is 2 Mbyte/sec, which is greater than
the speed of our Ethernet by a factor of two. Furthermore to build up our 
communication practically no time is needed,  while the build-up of the 
Ethernet connection is rather slow.
Even more 16 pairs of machines can
simultaneously communicate. Altogether we get two orders of magnitude 
better performance. Note that the system is scalable.
One can build machines with larger number of nodes. The
total inter-node communication performance is proportional to the
number of nodes.

The whole computer is rack mounted. This way the flat cables needed for 
communication are relatively short. At current communication speeds
this is not necessary. However, for future developments when communication
gets much faster this can be crucial.

\section{Description of  PMS Communication Hardware (CH) }

There are two communication cards in each machine. The CPU card contains
the main
circuits needed for transmitting data, while the Relay card contains the
connectors for the flat cables connecting the adjacent nodes and some
additional circuits.
\begin{figure}
  \centerline{\epsfig{file=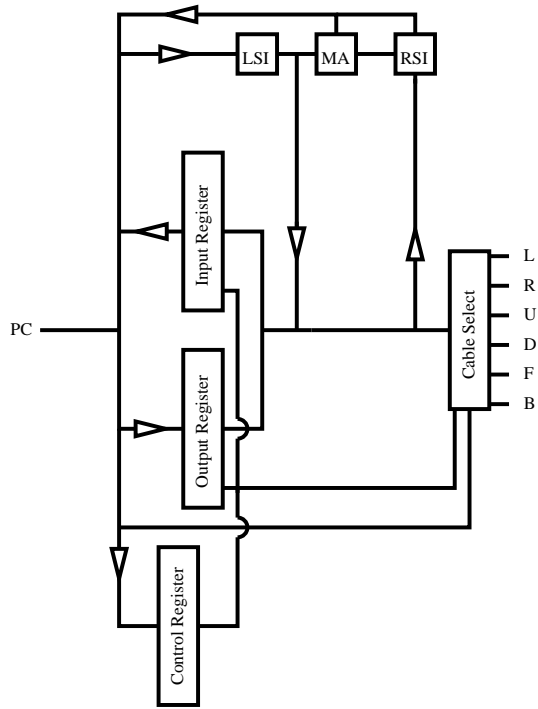,width=7cm}}
  \caption[h]{The PMS communication card \label{card_scem}}
\end{figure}
The block diagram of the cards is shown in Fig. \ref{card_scem}. 
The circuits of both
cards are included. There are two 16-bit buffers,
the output buffer and the input buffer, which accept the data coming from the
computer and from one of the adjacent nodes, respectively. The Control
Register
is used --among other things-- to clear the buffers and set the node to
either
sender or receiver state.
The Cable Select
circuit selects which direction the data is sent to or received from. The six
directions are labelled as left (L), right (R), up (U), down (D), front (F),
back (B).
If the same physical cable is selected by two adjacent nodes, one of them set
as sender
and the other as receiver, a physical connection is established and the
content of the output register of the sender is
immediately transferred to the receiver's input register.
The Local State Indicator (LSI) and the Remote State Indicator (RSI) are two
registers
to indicate the states of the nodes. There are 12 LSI and 12 RSI lines. They
correspond
to sending to and receiving from the six directions.
Each node can indicate its request for sending or receiving through the LSI
lines.
The RSI lines are identical to the six neighbours' LSI lines.
If there is a match between RSI and LSI signals (i.e. a send and a
corresponding receive request coincide) then the Match Any (MA) bit is set
and an IRQ
is generated on the PC bus. The interrupt is generated on both nodes at the
same time, so
the interrupt handlers on both machines can safely start transferring data
without any extra syncronization. The data can be transferred either 
by DMA or by consecutive I/O operations.

The Cable Select circuit, the LSI, RSI and MA registers together with the
flat
cable connectors are located on the relay card, while all other circuits are
located on the CPU card.

\section{Software}

As  mentioned above, the whole cluster is connected via an Ethernet
network. Each node has a special hostname that corresponds
to its location in the cluster. 
The hostnames
are s000,s001,s002,s003,s010 $\dots$ s133. The numbers in the 
hostname correspond to the coordinates of the node in the
three dimensional mesh.
When it is necessary for a node to identify itself (e.g. write/read priorities
during inter-node communication, see below) the
file '/etc/hostname' is used. The Ethernet network is used only 
for job management, i.e. to distribute
and collect data to and from the nodes. It is not used for 
inter-node communication during simulations. This is achieved 
by the special hardware described in the previous section.

In order to take advantage of the fast communication from applications
(e.g. high level C, C++ or Fortran code),
a low-level Linux kernel driver has been developed to access all the registers
of the communication card. From the user level cards can simply be reached by
reading or writing the device files '/dev/pms0, /dev/pms1 ... /dev/pms5'.
The six device files correspond to the six directions. A write operation
to one of these device files will         
transmit data to the corresponding direction, and reading from these device 
files reads
out previously transferred data. All the necessary input/output operations for
transferring data blocks are performed by the device driver. Notice the
important feature that this can be reached from any high level 
C, C++, Fortran code for which compilers are available.

The main structure of the device driver is similar to that of the card.
There are six read buffers, one for each direction in the main memory of the
machine and there is one write buffer. The data are always written to the write
buffer and read out from one of the read buffers.

The driver has two main parts. The first part is accessed from applications when
the user writes or reads any of the device files '/dev/pms*'. The other part is
the interrupt handler where the real data transfer takes place.

Whenever data are written to one of the device files, all the driver does is to
copy the written data to the write buffer and set the corresponding LSI send
signal to
indicate that a data send is requested. If the buffer is already full,
an error byte is returned to the application. Reading from the device files is
similar: if there are data in the corresponding read buffer, they are sent to the
application and the LSI receive line is set, since the node is ready to receive
new data. If the read buffer is empty, an 'End Of File' byte is returned to the
application.

When there is a coincidence between corresponding LSI and RSI signals,
an interrupt is generated by the card, which invokes the driver's
interrupt handler. It is the task of the handler to transfer data from the
sender to the receiver.
The interrupt handlers on the two communicating nodes start almost at
the same time. The difference may only be a few clock cycles. There is,
however, a need for synchronization. If the machines are ready to send
or receive the first byte they indicate it with their LSI lines. Notice 
that this will not cause an extra interrupt since interrupts are
disabled within the interrupt handler.
The sender first transmits the size of the package that will
follow in 16-bit words.
In the present version this is a 16-bit value, so the maximum size of a
package that can be transferred is 128 kbytes. Then the given
number of 16-bit values follow. Each word from the sender's write buffer
is copied to the receiver's read buffer. Finally, a 32-bit checksum is sent.
The receiver computes its own checksum and if it does not match the
received checksum, it is indicated to the sender and the whole transfer
is repeated. The final step in the interrupt handler is to clear the 
LSI lines of both nodes. On the one hand this indicates for the sender that the data
have been transferred and
the write buffer is empty again. On the other hand this tells the
receiver that the corresponding read buffer is full, so no new data can
arrive unless the buffer is emptied.

The buffer sizes are set in the driver to constant values. From the
previous paragraph it is clear that the maximum reasonable buffer size
is 128 kbytes. In order to save memory, while allowing large packages
at the same time the buffer size is set to only 64 kbytes at present.

The driver makes application programming quite easy. Communication can
be achieved by accessing the above mentioned device files. 
However, some C functions
have also been written to make writing applications even
simpler. These functions are the following:

{\it pms\_open} is used to initialize the card. It clears all
buffers, sets the LSI receive lines, clears the LSI send lines and
enables interrupts. The node thus becomes ready to receive data from any of
the neighbours.

{\it pms\_close} is used to close the card. All LSI lines
are cleared and interrupts are disabled. No further
communication may take place after this function call.

{\it pms\_send, pms\_recv}  are used to send and
receive data. Their parameters are the direction, the number of bytes
to send, and a pointer to the beginning of the data. On success they
return a positive value, otherwise a negative one. If there is no data
in the read buffer, {\it pms\_recv} returns 0.

The driver does not take care of any priority problems. It is possible
to write applications that will not work since all nodes are waiting
for data while none of them is sending anything. This is often the case
when the same code is running on all the nodes without any priority
check. There is a simple solution to these kind of problems. The
parity of the node is simply the
parity of the sum of the three digits in its hostname. Each time when
communication is performed, even nodes send data first and receive
afterwards, while odd nodes receive first and send their data afterwards.
This simple method works in most cases. Note that the code for it 
is located in the application (C, C++ or Fortran) and not in the driver.

\section{Performance}

We carried out lattice gauge theory simulations on our parallel
computer PMS. We used double precision variables.  
Two types of theories were studied.\\
a. We analyzed the pure SU(3) gauge theory with the 
simplest Wilson action. Overrelaxation and heatbath 
updating algorithms were used for the link variables.
The most CPU time consuming part, manipulation with
$3\times 3$ matrices, was written in assembly language.
This increased the speed of the code by more than a factor 
of two.\\
b. We studied the bosonic sector of the minimal
supersymmetric extension of the standard model (MSSM). 
SU(2) and SU(3) gauge fields were included with
two complex Higgs doublets and with left and
right handed coloured scalar quarks. Again, 
overrelaxation and heatbath  updating algorithms were 
used for both gauge link variables and scalar site variables.
The microcanonical overrelaxation update for scalar quarks 
is quite complicated and new, its details  will be
discussed elsewhere \cite{mssm_ewpt_prd}.
Manipulation with $3\times 3$ and $2 \times 2$ matrices 
was written in assembly language, which again increased 
the speed of the code essentially.

Having carried out lattice simulations for these two theories we 
obtained similar results for the speed of the
code and for the communication between nodes. 
The MSSM results for communication are actually somewhat better. The reason
for that is quite simple. The number of variables in the MSSM
is  larger by a factor of two than in pure SU(3) gauge theory; 
however, the number of floating point operations 
needed for a full update is more than an order of magnitude larger.
Thus, for the same lattice size the time needed to
transfer the surface variables --done by the
communication cards-- compared to the update time
is smaller for MSSM than for the pure SU(3) theory. 
We expect that a  similar though less pronounced effect 
should be observed for fermionic systems e.g. 
with multiboson \cite{multiboson}
algorithms. 
Since it is much more straightforward to compare our SU(3) results 
than those of the MSSM simulations with 
the results of the literature, 
in the following paragraphs we discuss the SU(3) case, too. 

For small lattice sizes the most economical way to use our
32 PC cluster is to put independent lattices on the different
nodes. The maximum lattice size in the SU(3) theory 
for 128 Mbyte memory is $\sim 20^4$,
or for finite temperature systems $6 \cdot 32^3$.
One thermalizes such a system on a single node, then distributes
the configuration to the other nodes and continues the
updating on all 32 nodes.
We measured the sustained performance of the cluster in this
case, which gave  $32 \times 152$Mflop=4.9Gflop. This
152Mflop/node performance means that one double precision operation is
carried out practically for every
third clock cycle of the 450 MHz,
whereas the nominal maximum of the 
processor is one operation for
every second clock cycle. As it was mentioned above, without assembly 
programming an approximate reduction factor of two  
in the performance was observed. 

Increasing the volume of the simulated system one can divide
the lattice between 2 nodes (the $2\times 4\times 4$ topology
has 2 nodes in one of the directions). For even larger lattices
one can use 4 nodes (4 in one direction), 8 nodes
($2\times 4$ in two directions) 16 nodes ($4\times 4$ in
two directions) or 32 nodes ($2 \times 4 \times 4$ in three
directions). Again, the most economical way to perform the simulations
is to prepare one thermalized configuration and put it on
other nodes (this method obviously can not be used for the
$2\times 4\times 4$ topology, because in this case the whole 
machine with 32 nodes is just one lattice). 

\begin{figure}
\centerline{\epsfig{file=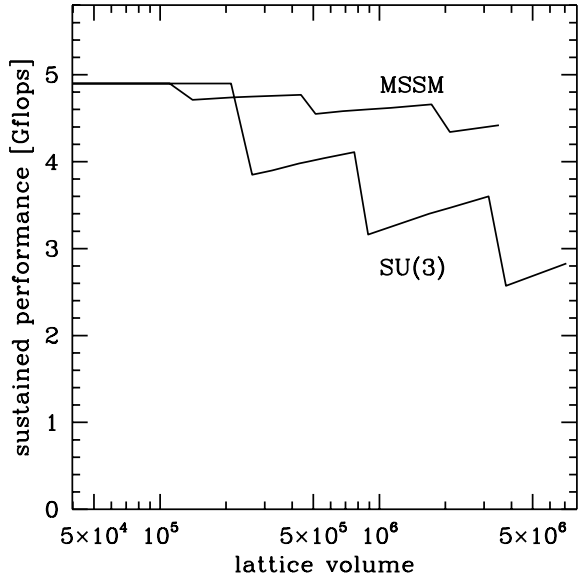,width=8cm}}
\caption[a]{Sustained performance of  PMS  as a function
of the lattice volume for pure SU(3) gauge theory and for MSSM. 
The endpoints of the lines correspond to the largest
volumes which can be simulated on a 32 PC cluster.
\label{perform}}
\end{figure}

Based on our measurements we determined the sustained performance
of a 32-node PMS cluster as a function of the lattice volume. The
result for a set of lattice volumes for finite temperature systems 
with temporal extension,
$L_t=6$ for SU(3) and $L_t=4$ for MSSM can be seen on Fig. 
\ref{perform}. Clearly, the largest volume one can reach is approximately
twice as large for the SU(3) gauge theory than for the MSSM. 
For both cases there are regions where the performance increases
with the volume. This can be easily explained by the fact that larger 
volume means better surface/volume ratio, thus better performance.
There are three drops in the performance for both SU(3) and MSSM. 
They correspond to
lattice volumes for which new communication directions were opened
(or, in other words, the dimension of the mesh of the nodes on which the lattice
was divided, increased by one) in order to fit the lattice into the 
available RAMs. As it can be seen the performance for the MSSM 
is still very high even at the largest volume with three-dimensional
communication: it is just 10\% smaller than the performance without
communication. This plot gives us the optimum architecture of
such a parallel computer. The number of nodes and the
number of communication directions used for a given lattice 
should be as small as possible simultaneously. 
This means  a $2\times 4\times 4$
topology for 32 nodes and a $2\times 4\times 8$ topology 64 nodes. 

Despite the fact that the speed of the communication between two nodes
is not that high (2 Mbit/sec) the performance of the cluster is quite good. The
reason for this is the high speed of the individual nodes (450 MHz) and the
large RAM on each node. This sort of design does not need a division of the
lattice to hundreds of sub-lattices, thus it does not need a very fast
communication. Clearly, the use of PCI communication instead of ISA
gives an order of magnitude faster communication, which increases the
potential of such a machine.

\begin{figure}
\centerline{\epsfig{file=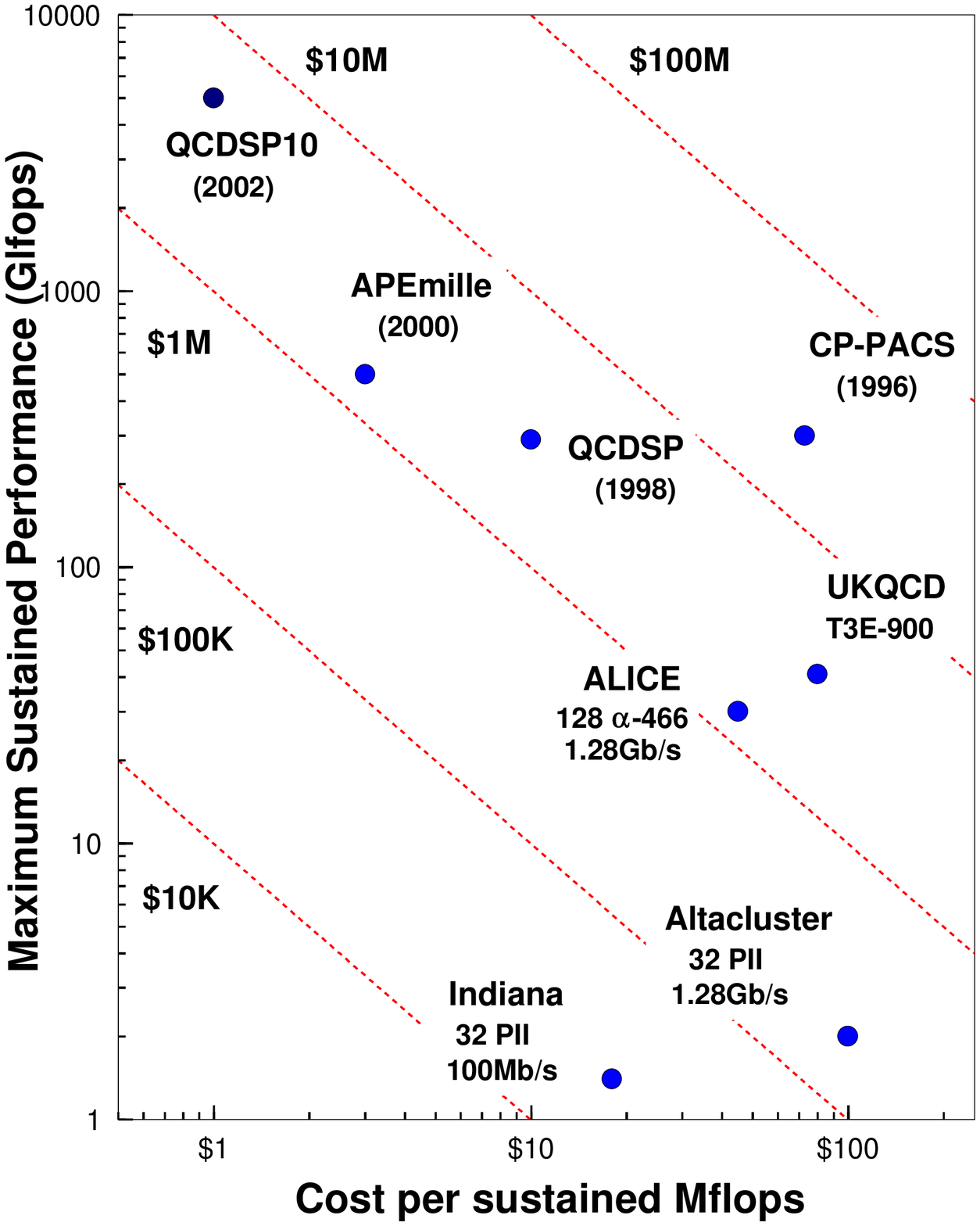,width=10cm,height=10cm}}
\vspace{-10cm}
\centerline{\epsfig{file=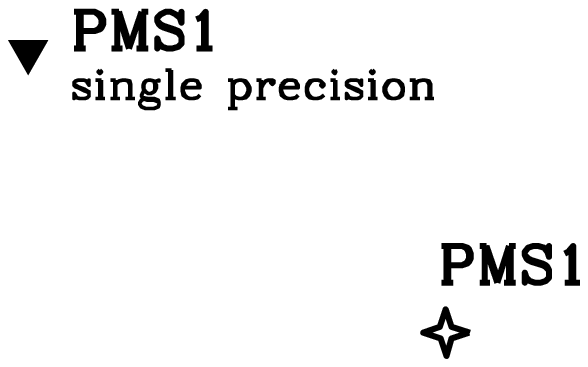,width=10cm}}
\caption[a]{Sustained performance of different machines
used in lattice gauge theory as a function of price/performance.
Our PMS machine is presented as a 32 PC based system.
\label{compare}}
\end{figure}

In order to compare our machine with other existing (and proposed) 
parallel systems
used in lattice gauge theory, we use the plot of N. Christ \cite{christ}.
As it can be seen in Fig. \ref{compare},
 PMS1 is perhaps the best existing machine  
as far as price/sustained performance is concerned. 
It is particularly important to emphasize 
 that the presented performance 3\$/Mflop with $\approx$4Gflop total 
 performance of  PMS1  is calculated
for double precision operations (while some of the performances of the 
machines on the plot are given for single precision operations). 
For single precision simulations one can use the MMX instruction set which is
8 times faster than the double precision operations (4 operations for each
clock cycle). We estimated the single precision performance by assuming that
MMX programming results in 20\% decrease in performance. The total performance
of PMS1 is $\approx$27Gflop with 0.45\$/Mflop price-to-sustained performance
ratio. (Notice that our measured performance is based on double precision
code, whereas single precision numbers are estimated ones.) 
Our PMS1 machine, similarly to other workstation farms has a moderate 
maximum sustained
performance as compared to Teraflop-scale machines (CP-PACS \cite{CP-PACS} or QCDSP \cite{QCDSP}). 
However, PMS1 has a  much better price/performance ratio than  other workstation
farms. The reason for this is threefold.\\
a. At the day of purchase our choice for the PC components was optimized 
for price/performance.\\
b. Machines with 100Mbit fast Ethernet (e.g. Indiana) are inexpensive, but
communication is slow, resulting in a high price/performance ratio. \\
c. Machines with much faster networks using Myrinet (e.g. Alice \cite{alpha} 
and Altacluster)
give higher performance. However, this is balanced in the price/performance
ratio by the higher cost of the network.\\
The essential feature of  PMS architecture is that the speed of the 
communication (depending on the size of the machine) is comparable to that of
Myrinet. However, the price for such a communication is as low as 
$\approx$40\$/node, which is more than an order of magnitude less than the
cost of Myrinet.

\section{Conclusion}
We have presented a description of the status of  PMS project. Details of 
the communication hardware and  software have been explained. The existing 
32-node machine (PMS1) is able to perform medium size LGT simulations in case 
of nearest neighbour interactions, distributing the lattice on several nodes, 
if necessary. This is done at a very good price/sustained performance ratio.
Problems requiring more number crunching and less communication give a  
better performance. 

We plan to increase the number of connected nodes and implement the PMS CH 
for faster buses in order to increase the maximum size of programs and enhance
communication speed.

Finally, we would like to emphasize that the design of our communication hardware and the
developed Linux drivers are freely available for non-profit organizations.
Those who intend to build a PMS architecture machine have simply to buy PCs 
produce the communication cards and connect the nodes by flat cables.
Furthermore they have to install Linux and our Linux driver.
Questions and requests for information should be sent to pms@labor.elte.hu.

\section{Acknowledgement}
This work was supported by Hungarian Science Foundation Grants under Contract
Nos. OTKA-T22929-T29803-F17310-M28413/FKFP-0128/1997.

\section{Appendix}

The following C code is a simple example to illustrate the usage of PMS
communication cards in a given application. This example transfers random numbers
in each direction. It can be used to test all channels. We performed
this test on all cards by sending 100 Gbytes of data on
each channel. We observed no errors.
The header file 'pms.h' that contains the function prototypes
is also included.\\

%/*****************  tst.c  *********************/ 
\baselineskip=-0.4truecm
\noindent

{\small
\begin{verbatim}
#include<stdio.h>
#include<stdlib.h>
#include<string.h>
#include"pms.h"               /* PMS function prototypes */
#define MUL 1579              /* Random number generator multiplier */
#define N 1000                /* Number of packages to transfer */
#define K 16384               /* Size of each package in words */
unsigned short j_r=1, j_s=1;  /* initial values for random number generators */

int send(int dir) {
static unsigned short buf[K];
int i,err;
   
  for (i=0; i<K; i++) {       /* generate a series of random numbers */
    j_s*=MUL;
    j_s++;
    buf[i]=j_s;
  }
  err=pms_send(buf,K,dir);    /* send the random numbers to direction 'dir' */
  return err;
} /* end of send() */
/*********************************************/

int recv(int dir) {
  static unsigned short buf[K];
  int i,err;

  while((err=pms_recv(buf,K,dir))==0); /* Wait for data: read until data comes */
  if (err>0) {
    err=0;
    for (i=0; i<K; i++) {     /* Generate the same random numbers */
      j_r*=MUL;               /* and check for errors */
      j_r++;
      if (buf[i]!=j_r) {
        err++;
        printf("Error #%d\n",i);
        printf("Sent: 0x%4.hX\tRecvd: 0x%4.hX\n",j_r,buf[i]);
      }
    }
  }
  return err;
}/* end of recv() */
/*********************************************/

void main(int argc, char * argv[]) {
int i;
FILE * hostname;
char host[10];
int pos[3],par;

  hostname=fopen("/etc/hostname","r"); /* Get the hostname of the current node */
  fscanf(hostname,"%s\n",host);
  pos[0]=host[1]-48;
  pos[1]=host[2]-48;
  pos[2]=host[3]-48;
  par=(pos[0]+pos[1]+pos[2])\&1; /* Get the parity of the node */
  printf("position: %d, %d, %d\n",pos[0], pos[1], pos[2]);
  printf("parity: %d\n",par);
  if (pms_open()<0) {         /* Initialize the card */
    printf("PMS card can not be opened\n");
    exit(-1);
  }
  for (i=0; i<N; i++) {
    printf("Testing 1D+ -> 1D-\n"); /* Test the first direction */
    if (!par) {  /* Priority check: even nodes send first, odd nodes */
                 /* receive first */
      if (send(0) <=0) printf("Send error\n");
      if (recv(1) !=0) printf("Recv error\n");
    } else {
      if (recv(1) !=0) printf("Recv error\n");
      if (send(0) <=0) printf("Send error\n");
    }

    printf("Testing 1D- -> 1D+\n"); /* Test the second direction */
    if (!par) {
      if (send(1) <=0) printf("Send error\n");
      if (recv(0) !=0) printf("Recv error\n");
    } else {
    if (recv(0) !=0) printf("Recv error\n");
    if (send(1) <=0) printf("Send error\n");
  }
  printf("Testing 2D+ -> 2D-\n"); /* Test the first direction */
  if (!par) {
    if (send(2) <=0) printf("Send error\n");
    if (recv(3) !=0) printf("Recv error\n");
  } else {
    if (recv(3) !=0) printf("Recv error\n");
    if (send(2) <=0) printf("Send error\n");
  }

  printf("Testing 2D- -> 2D+\n"); /* Test the second direction */
  if (!par) {
    if (send(3) <=0) printf("Send error\n");
    if (recv(2) !=0) printf("Recv error\n");
  } else {
    if (recv(2) !=0) printf("Recv error\n");
    if (send(3) <=0) printf("Send error\n");
  }
  printf("Testing 3D+ -> 3D-\n"); /* Test the first direction */
  if (!par) {
    if (send(4) <=0) printf("Send error\n");
    if (recv(5) !=0) printf("Recv error\n");
  } else {
    if (recv(5) !=0) printf("Recv error\n");
    if (send(4) <=0) printf("Send error\n");
  }

  printf("Testing 3D- -> 3D+\n"); /* Test the second direction */
  if (!par) {
    if (send(5) <=0) printf("Send error\n");
    if (recv(4) !=0) printf("Recv error\n");
  } else {
    if (recv(4) !=0) printf("Recv error\n");
    if (send(5) <=0) printf("Send error\n");
  }
  pms_close(); /* Close the card */
}/* end of main() */
/*********************************************/

(file: pms.h)

extern int mynode;
extern int node_par;

int pms_open(void);
void pms_close(void);
int pms_send(void * buf, int count, int dir);
int pms_recv(void * buf, int count, int dir);


\end{verbatim}

\end{document}